\documentclass[twocolumn,pra,aps]{revtex4}

\usepackage{graphicx}
\usepackage{bm}
\usepackage{amsmath}
\usepackage{amssymb}
\usepackage{underscore}
\usepackage{color}
\usepackage{amsthm}
\usepackage{ulem}

\def\squareforqed{\hbox{\rlap{$\sqcap$}$\sqcup$}}
\def\qed{\ifmmode\squareforqed\else{\unskip\nobreak\hfil
\penalty50\hskip1em\null\nobreak\hfil\squareforqed
\parfillskip=0pt\finalhyphendemerits=0\endgraf}\fi}

\def\duzomniejsze{<\kern-.7mm<}
\def\duzowieksze{>\kern-.7mm>}

\def\textbf#1{{\bf #1}}
\def\beq{\begin{equation}}
\def\eeq{\end{equation}}
\def\be{\begin{equation}}
\def\ee{\end{equation}}
\def\bal{\begin{align}}
\def\eal{\end{align}}
\def\ben{\begin{eqnarray}}
\def\een{\end{eqnarray}}
\def\beqa{\begin{eqnarray}}
\def\eeqa{\end{eqnarray}}
\def\eea{\end{array}}
\def\bea{\begin{array}}

\newcommand{\bei}{\begin{itemize}}
\newcommand{\eei}{\end{itemize}}
\newcommand{\bee}{\begin{enumerate}}
\newcommand{\eee}{\end{enumerate}}
\newcommand{\nc}{\newcommand}
\nc{\1}{{\openone}}

\def\ccal{{\cal C}}

\def\>{\rangle}
\def\<{\langle}

\def\ot{\otimes}

\newtheorem{lemma}{Lemma}
\newtheorem{corollary}{Corollary}
\newtheorem{proposition}{Proposition}
\newtheorem{theorem}{Theorem}
\newtheorem{definition}{Definition}
\newtheorem{observation}{Observation}

\newtheorem{rem}{Remark}

\theoremstyle{plain}
\newtheorem*{corone}{Corollary 1}
\newtheorem*{prop1}{Proposition 1}
\newtheorem*{thm2}{Theorem 2}
\newtheorem*{thm3}{Theorem 3}

\def\bed{\begin{definition}}
\def\eed{\end{definition}}
\def\bel{\begin{lemma}}
\def\eel{\end{lemma}}
\def\bet{\begin{theorem}}
\def\eet{\end{theorem}}

\newcommand{\ketbra}[2]{\left|#1\middle\rangle\middle\langle#2\right|}
\newcommand{\de}[1]{\left( #1 \right)}

\newcommand{\DE}[1]{\left\{#1\right\}}

\newcommand{\ie}{\textit{i.e. }}

\begin{document}

\title{Bounds on quantum nonlocality via partial transposition}

\begin{abstract}
We explore the link between two concepts: the level of violation of a Bell inequality by a quantum state and discrimination between two
states by means of 
restricted classes of operations, such as local operations and classical communication (LOCC) and separable ones.
For any bipartite Bell inequality, we show that 
 its value on a given quantum state cannot exceed the classical bound  by more
than 
the maximal quantum violation shrunk by a factor related to distinguishability of this state
from the separable set by means of 
some restricted class of operations. 
We then consider the general scenarios where the parties are allowed to perform a local pre-processing of 
many copies of the state before the Bell test (asymptotic and hidden-nonlocality scenarios). We define the asymptotic relative entropy of nonlocality and, for PPT states,
we bound this quantity by the relative entropy of entanglement of 
the partially transposed state.
The bounds are strong enough to limit the use of certain states
containing private key in the device-independent scenario.
\end{abstract}

\author{Karol Horodecki$^{1,2}$, Gl\'aucia Murta$^{2,3}$ }
\affiliation{$^1$Institute of Informatics, University of Gda\'nsk, 80--952 Gda\'nsk, Poland}
\affiliation{$^2$National Quantum Information Centre in Gda\'nsk, 81--824 Sopot, Poland}
\affiliation{$^3$Departamento de F\'isica, Universidade Federal de Minas Gerais,
 Caixa Postal 702, 30123-970, Belo Horizonte, MG, Brazil}

\maketitle

\bibliographystyle{apsrev}

Nonlocality is one of the most interesting phenomena emerging from quantum multipartite states, extensively studied in the recent years
\cite{review-nonloc}. 
The quantitative study of nonlocality has two different approaches, one is to ask, for a fixed Bell scenario, what is the best one can obtain optimizing over all possible quantum resources 
(states and measurements) \cite{Tsirelson, navascues-2008-10, Palazuelos2010, Palazuelos2011,Palazuelos2012,Palazuelos3part}. 
A converse approach is to ask for a fixed quantum state, or a class of states, what is the best one can obtain using this state as a resource, 
\ie optimizing over all Bell scenarios. 
Some references in this second approach include the seminal work of Werner \cite{Werner1989} exhibiting a local model for 
projective measurements for $U\otimes U$-invariant states (see also \cite{Barrett-local}). Another result
showing that typically the violation of correlation Bell inequalities by multipartite qudit states is very small \cite{Drumond2012}. 
And an hierarchy of semidefinite programs that allows one to bound the violation achievable by PPT states \cite{hierarchyPPT}. 

Here we follow the second approach, with the aim to show that certain states, despite being entangled, exhibit very limited gain of nonlocality.
To achieve this, we base on the concept of state discrimination by means of local operations and classical communication (LOCC) \cite{Bennett-nlwe},
a subject extensively studied in the last decade (see e.g. \cite{Duan-discrim} for recent results, and references therein).
It has been shown, that there exist pairs of states which are hardly distinguishable from each other by means of LOCC, 
although being almost orthogonal, i.e. almost perfectly distinguishable by means of global operations \cite{Bennett-nlwe,hiding-ieee,hiding-prl,WernerHide}. 
In \cite{karol-PhD} it is shown that there exist even entangled states containing bit of privacy, which are almost indistinguishable 
by LOCC operations from some separable 
(insecure) states. This fact has been shown recently to rule them out as a potential resource for swapping of a private key, 
in the so called quantum key repeaters \cite{BCHW-swapping}. 

We base also on the idea stated in \cite{Hyllus-Bell-wit,Terhal-Bell-wit}, where the Bell inequality is considered as a particular witness of entanglement.
The link between quantum nonlocality and state discrimination that we start from,
amounts to a simple observation that if a given state is hardly distinguishable from some separable one, it can not exhibit
large violation in any Bell scenario, or else, one could use the  
procedure of checking the violation of a Bell inequality to discriminate between 
these two states (a quantitative version of this fact has been derived in \cite{BrunnerVertesiTh1}).
Here we refine these ideas,  using partial transposition to explore the fact that Bell
inequalities are implemented 
by a much smaller class of operations, the local ones.
We obtain non-trivial upper bounds on the amount of quantum nonlocality both in the single copy case for arbitrary 
bipartite states, as well as in the asymptotic and hidden-nonlocality scenarios for states with positive partial transpose. To the 
best of our knowledge, this is the first quantitative approach for the latter two scenarios.

We 
start by deriving  bounds on the violation of a Bell inequality on a single copy of a quantum state $\rho$. 
It exceeds the classical value by the maximal quantum value
shrunk by a factor related to distinguishability between the state and separable states by means of separable operations 
\footnote{Separable operations are the quantum operations which can be written in the form $\Lambda_{sep}(\rho)= \sum_i M_i \rho M^{\dagger}_i$ 
s.t. $M_i=A_i\otimes B_i$, which include 
the LOCC operations. They are a subset of a larger class called $ppt$ operations.}. 
Since it is hard to compute distinguishability by separable operations,
following \cite{WernerHide, RainsPPToperations}, we bound 
it using partial transposition. The bounds are strong enough to guarantee very limited nonlocality for some entangled states
which contain secure quantum key. For this reason the use of these states for  device independent security 
appears to be limited.
We then generalize the bounds to the asymptotic case of a large number of copies of the
state, showing that for PPT states the \textit{asymptotic relative entropy of nonlocality} \cite{vDamGrunwaldGill} 
is upper bounded by the relative entropy of entanglement of 
the partially transposed state, $\rho^{\Gamma}$. We also consider a hidden-nonlocality scenario, and we prove the same upper bound 
for the \textit{asymptotic relative entropy of hidden-nonlocality}, which we define here. 
Interestingly, to achieve the results, we apply techniques of \cite{BCHW-swapping}, which were developed for the quantum key repeaters scenario.
Details of the  proofs of the above results can be found in the Supplemental Material.

{\it Notation.---}  By bipartite box we refer to the conditional probability distribution of outputs $a$ and $b$ of Alice and Bob
given inputs $x$ and $y$ respectively, $P(ab|xy)$. 
In what follows, by the set $\mathcal{S}\equiv\{s^{a,b}_{x,y}\}$ we denote the coefficients of a particular Bell inequality, so that
$\sum_{a,b,x,y}s^{a,b}_{x,y}P(ab|xy)$ is the value of the Bell inequality $\mathcal{S}$ on a particular box $P(ab|xy)$.
Denoting the 
maximal value of the Bell expression $\mathcal{S}$ over all boxes 
in classical case $C(\mathcal{S})$, quantum $Q(\mathcal{S})$ and supraquantum $NS(\mathcal{S})$, 
we have, {withou loss of generality, the following relation: $C(\mathcal{S}) \leq Q(\mathcal{S}) \leq NS(\mathcal{S})$.
For a bipartite
state $\rho_{AB}$ and the set of POVMs $\DE{M_{xy}}$, $M_{xy}= \DE{A_{a|x}\otimes B_{b|y}}$, we represent the corresponding box by $\{Tr M_{xy} \rho_{AB}\}$,
and denote the value of the Bell expression $\mathcal{S}$ for these particular POVMs by  $\mathbf{S}(\rho_{AB})$, i.e. $\mathbf{S}(\rho_{AB})= Tr \mathbf{S}\rho_{AB}$ 
where $\mathbf{S} = \sum_{a,b,x,y} s^{a,b}_{x,y} A_{a|x}\otimes B_{b|y}$.

{\it Bounds for Bell inequalities.---} We are ready to relate the value of a Bell inequality on two bipartite quantum states to the bound on their
distinguishability by means of PPT operations.

{\theorem Given two bipartite states $\rho,\sigma \in B(\ccal^{d}\otimes\ccal^{d})$, a Bell inequality $\{s^{a,b}_{x,y}\}$
and a set of quantum POVMs $\{A_{a|x}\otimes B_{b|y}\}$, it holds that:
\be
|\mathbf{S}(\rho) - \mathbf{S}(\sigma)| \leq ||\mathbf{S}^{\Gamma}||_{\infty} ||\rho^{\Gamma} - \sigma^{\Gamma}||.
\label{eq:gen-discrim}
\ee\label{thm:2-state-disc}where $||.||$ denotes the trace norm, $||X||_{\infty}$ is the largest eigenvalue in modulus of operator $X$, and  $\Gamma$ denotes partial transposition.
\label{thm:ppt-norm-bound}
}

Before proving Theorem \ref{thm:ppt-norm-bound}, let us note that $\mathbf{S^{\Gamma}}$ is
also a Bell operator since partial transposition maps $\{A_{a|x}\otimes B_{b|y}\}$ into another set of allowed measurements $\{A_{a|x}\otimes B_{b|y}^T\}$, 
and that $||\mathbf{S^{\Gamma}}||_{\infty}$ is nothing but the largest quantum value of the Bell expression given particular 
measurements $\{A_{a|x}\otimes B_{b|y}^T\}$.
The second term of RHS upper bounds the distinguishability
of these two states by separable operations \cite{RainsPPToperations,WernerHide}. 
A weaker form of Theorem \ref{thm:ppt-norm-bound}, relating the Bell value with distinguishability under
general measurements, was similarly derived in \cite{BrunnerVertesiTh1}.

{\it Proof}. We start by considering Bell inequalities with positive coefficient, \ie $s^{ab}_{xy} \geq 0 \; \forall \, a,b,x,y$
(note that using the normalization condition we can rewrite any Bell inequality with positive coefficients only).
We show the sequence of (in)equalities and comment it below:
\begin{align}
\mathbf{S}(\rho) - \mathbf{S}(\sigma) = & \sum_{a,b,x,y} Tr s^{a,b}_{x,y} A_{a|x}\otimes B_{b|y} (\rho - \sigma)  \nonumber \\
 =&\sum_{a,b,x,y} Tr s^{a,b}_{x,y} A_{a|x}\otimes (B_{b|y})^T (\rho - \sigma)^{\Gamma}  \nonumber\\
=& ||\mathbf{S}^{\Gamma}||_{\infty} Tr {\mathbf{S}^{\Gamma}\over ||\mathbf{S}^{\Gamma}||_{\infty}}(\rho^{\Gamma} -\sigma^{\Gamma})  \label{eq:infty-norm}\\
\leq & ||\mathbf{S}^{\Gamma}||_{\infty} \sup_{M\geq 0, M\leq \1} Tr M(\rho^{\Gamma} -\sigma^{\Gamma})  \nonumber \\
= &||\mathbf{S}^{\Gamma}||_{\infty}||\rho^{\Gamma} - \label{eq:positive} \sigma^{\Gamma}||.\nonumber
\end{align}
In the first equality we use the linearity of trace function, then the well known identity $Tr XY = TrX^{\Gamma}Y^{\Gamma}$.
In the fourth step, we use the fact that Bell inequality is non-negative, i.e. $s^{a,b}_{x,y} \geq 0$, so the operator $\mathbf{S^{\Gamma}}$ is positive 
as a sum of positive operators $A_{a|x}\otimes B_{b|y}^T$ (note that transposition does not change positivity).
Further, by definition of infinity norm, we have ${\mathbf{S}^{\Gamma}\over ||\mathbf{S}^{\Gamma}||_{\infty} }\leq \1$. 
Finally we use the fact that trace norm is the supremum over positive operators less than identity (see \cite{NielsenChuang}), and the assertion follows.

For general Bell inequalities, where not all coefficients are positive, we can derive Theorem 1 using H\"{o}lder inequalities for $p-$norms (see Supplemental
Material).
\qed

{\rem Let us note here that, in eq. (\ref{eq:infty-norm}), for Bell inequalities written with positive coefficients, 
we can have a much tighter bound: $|\mathbf{S}(\rho) - \mathbf{S}(\sigma)| \leq ||\mathbf{S}^{\Gamma}||_{\infty} ||\rho^{\Gamma} -\sigma^{\Gamma}||_{sep}$,
where $||X||_{sep}\equiv \sup_{M \in sep, M\leq \1}Tr M X$, where $ M = \sum_i M^A_i\otimes M^B_i$ for $M^A_i$ and $M^B_i$ being positive operators. 
It is however hard to evaluate the latter quantity, hence we focus here on the upper bound on it.
\label{rem:sep-norm}}
 
From Theorem \ref{thm:2-state-disc} we have an immediate corollary related to the fact that separable states, i.e. states of the form 
$\sigma_{AB} = \sum_{i} p_{i} \sigma_A^i\otimes \sigma_B^i$, yield boxes that have local hidden variable model. In consequence, 
$\mathbf{S}(\sigma_{AB}) \leq C(\mathcal{S}) \;\; \forall \; \sigma_{AB} \in SEP$, with $SEP$ denoting the set of separable states.

Also, let us denote 
\be
Q_{\mathcal{S}}(\rho) := \sup_{\DE{A_{a|x}\otimes B_{b|y}}} \sum_{a,b,x,y}s^{a,b}_{x,y}Tr A_{a|x}\otimes B_{b|y} \rho,
\ee
with supremum taken over all POVM elements $\{A_{a|x}\otimes B_{b|y}\}$.  
Note that $Q(\mathcal{S}) = \sup_{\rho}Q_{\mathcal{S}}(\rho)$, and it is straightforward that  $||\mathbf{S}^{\Gamma}||_{\infty}$ is upper
bounded by $Q(\mathcal{S})$. Hence we have:
{\corollary  For any bipartite Bell expression $\mathcal{S}$, and state $\rho$, it holds that:
\be
Q_{\mathcal{S}}(\rho) \leq C(\mathcal{S}) + Q(\mathcal{S})\inf_{\sigma \in SEP} ||\rho^{\Gamma}  - \sigma^{\Gamma}|| .
\label{eq:C-Q-bound}
\ee
\label{cor:C-Q-bound}
}

Note that one can drop partial transposition of $\sigma$ in \eqref{eq:C-Q-bound} since we take infimum over all separable states. 

Corollary \ref{cor:C-Q-bound} shows that, given a Bell inequality $\mathcal{S}$, the best quantum violation one can achieve with a particular state $\rho$ 
(optimazing over all possible measurements) cannot outperform the classical bound by the Tsirelson bound of the Bell inequality shrunk 
by a factor reporting the distinguishability of state $\rho$ from the set of separable states by means of a restricted class of operations.

One can expect that the same bound should hold for all states which are to the same extent indistinguishable from
separable ones.  To this end we introduce the hierarchy of sets $D(\epsilon)$: 
\be \label{D-epsilon}
D(\epsilon) : = \{ \rho : \exists_{\sigma \in SEP} \,\, ||\rho^{\Gamma} - \sigma^{\Gamma}||\leq \epsilon\}.
\ee
Observe that $D(\epsilon)$ is a convex set, which includes $SEP$ for any $\epsilon >0$.
In consequence, due to Corollary \ref{cor:C-Q-bound}, we have the following dependence (see Fig. \ref{fig:d-epsilon}):
{\corollary For any bipartite Bell expression $\mathcal{S}$ and $\epsilon >0$, it holds that :
\be 
\sup_{\rho \in D(\epsilon)} Q_{\mathcal{S}}(\rho) \leq C(\mathcal{S}) + \epsilon Q(\mathcal{S}).
\ee}

Note here, that according to Remark \ref{rem:sep-norm}, a stronger version of the above result, for positive coefficients Bell inequalities,
holds for the set $D(\epsilon)$ 
defined with respect to 
$||.||_{sep}$ norm instead of the norm based on partial transposition.

\begin{figure}	
	\centering
	\includegraphics[scale=0.42]{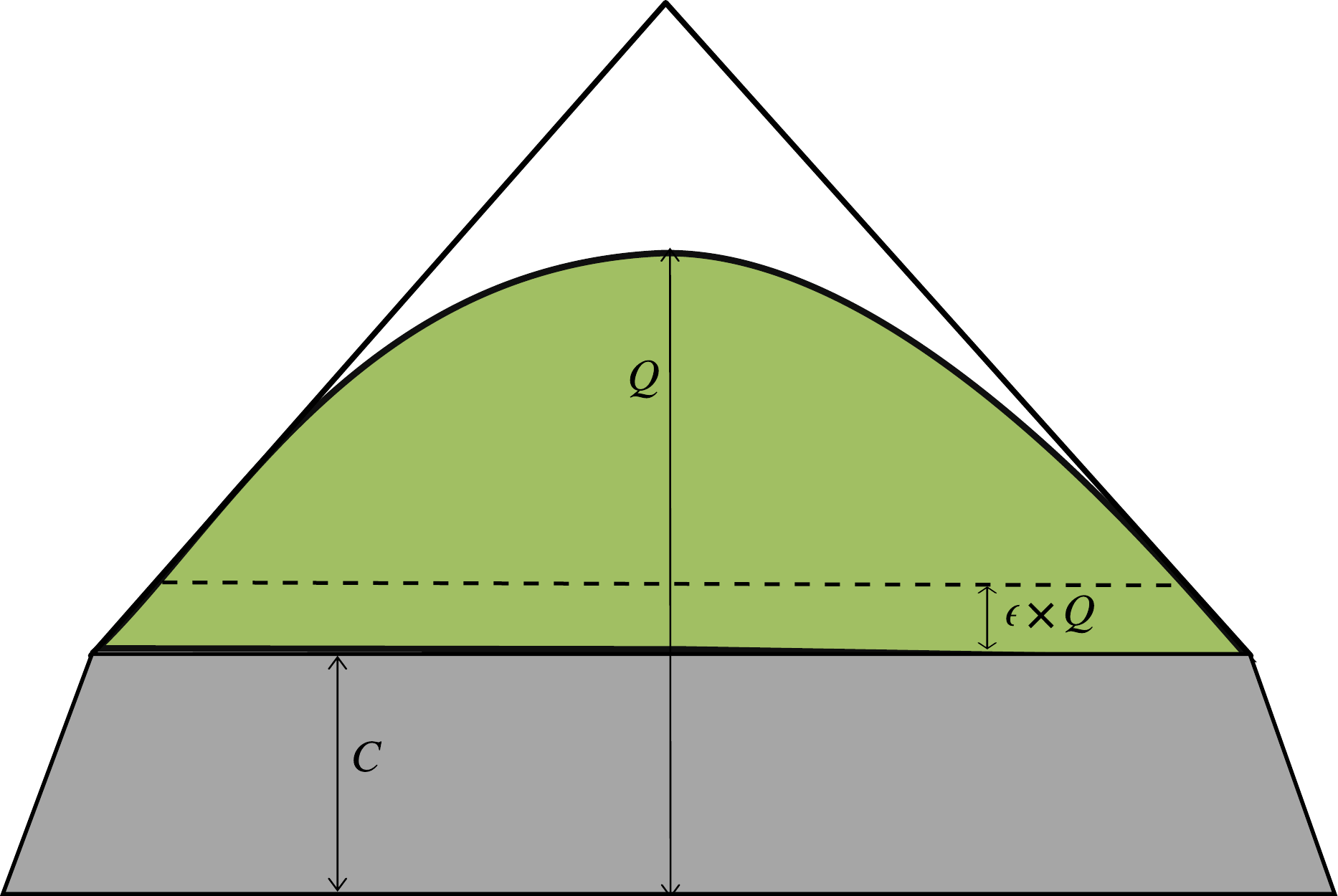}
	\caption{For states in $D(\epsilon)$, the set of states whose partial transposition is at most $\epsilon$ far from some separable state (see 
	eq. \eqref{D-epsilon}), the 
	violation of a Bell inequality is limited by the maximum quantum value, $Q$, shrunk by $\epsilon$ (dashed line).
		}
		\label{fig:d-epsilon}
\end{figure}

{\it Examples.---} Basing on \cite{karol-PhD} and \cite{BCHW-swapping} we exhibit some examples of entangled states that
have negligible violation. In our construction we base on private bits \cite{pptkey,keyhuge}.
A private bit can be described as follows: 
\ben
\gamma_{X}={1\over 2}[|00\>\<00|\otimes \sqrt{XX^{\dagger}}+ |00\>\<11|\otimes X +  \nonumber\\
|11\>\<00|\otimes X^{\dagger} +  |11\>\<11|\otimes\sqrt{X^{\dagger}X} ] .
\een
where X is an arbitrary operator with trace norm 1.
We then see that it resembles a singlet state with ``operator amplitudes'' which are functions of X.

Consider a private state 
defined by $X = {1\over d^2}\sum_{i,j=0}^{d-1} |ij\>\<ji|$ being the (normalized) swap operator. Then for the CHSH inequality we have the following bound:
\be
Q_{CHSH}(\gamma_{X}) \leq 2 + {\sqrt{2} + 1\over 2\sqrt{2}{d}}.
\label{eq:1-example}
\ee

In \cite{PrivateNonlocality} it was shown that all perfect private states exhibit nonlocality.
With our techniques we are able to find non-trivial upper bounds for exact private states as well as for PPT approximate private states.

Consider the following PPT state:
\begin{align}
\rho_{p} =(1-p) \gamma_{X} + {p\over 2}[|01\>\<01|\otimes \sqrt{YY^{\dagger}} + |10\>\<10|\otimes \sqrt{Y^{\dagger}Y}] 
\label{eq:ppt-1example}
\end{align}
with $X= {1\over d_s \sqrt{d_s}} \sum_{i,j=0}^{d_s-1} u_{ij}|ij\>\<ji|$ and $Y = \sqrt{d_s}X^{\Gamma}$, and $u_{ij}$ are 
elements of an unitary matrix with all $|u_{ij}| = {1\over d_s}$ (an example is the quantum Fourier transform), and $p={1\over \sqrt{d_s}+1}$. 
By Corollary \ref{cor:C-Q-bound}, we have
\begin{align}
Q_{\mathcal{S}}(\rho_{p}) \leq C(\mathcal{S}) + Q(\mathcal{S}) \frac{1}{\sqrt{d_s}}.
\label{eq:ppt-example}
\end{align}

Following \cite{BCHW-swapping}, we can also obtain a bound for states which are (up to local unitary transformation) invariant under
partial transposition, and at the same time entangled and containing private key. The states considered in the proposition below are defined in \cite{pptkey}
(see Supplemental Material).

{\proposition There exist bipartite states $\rho \in B(\ccal^2\otimes\ccal^2 \otimes (\ccal^{d^k}\otimes\ccal^{d^k})^{\otimes m})$ with $d=m^2$, $k=m$ satisfying
$K_D(\rho^{\Gamma}\otimes\rho) \rightarrow 1$ with increasing $m$, such that: 
\be
Q_{\mathcal{S}}(\rho\otimes\rho^{\Gamma})\leq C(\mathcal{S}) +{ Q(\mathcal{S})\over 2^{m-1}}.
\label{eq:prop-example}
\ee
\label{prop:example}
}

Proposition \ref{prop:example} shows that for some class of states invariant under partial transposition, although the rate of  distillable key can be made 
arbitrarily close to 1 by increasing the dimension of the systems, the possibility of violating any Bell inequality is bounded by an amount
vanishing  
with the dimension of the system.

In \cite{Palazuelos2011} it is shown that a bipartite Bell inequality with $n$ inputs and $k$ outputs satisfies
$ Q(\mathcal{S}) \leq C(\mathcal{S}) \min \DE{n,k}$,
up to some universal constant independent of the parameters of the scenario. 
With this we see that the bound \eqref{eq:C-Q-bound} ensures that for any fixed Bell scenario, as we wish to increase the key rate obtained from the exhibited families of states,  
the possibility of observing a violation of a Bell inequality vanishes.

{\it Bound on asymptotic nonlocality.---} In considerations above, we have provided bounds which hold for single copy of a quantum state. 
However, in case of the first example,
the state $\rho$ is distillable \cite{AH-pditdist}, hence, as it was noted by Peres \cite{Peres1996}, a pre-processing of many copies of a state by 
local operations, before the Bell test, could lead to the violation of a Bell inequality, even for states that have local model for the single copy level.
Here we quantify the asymptotic nonlocality by defining the  
asymptotic relative entropy of nonlocality 
(see \cite{vDamGrunwaldGill}) and applying methods of \cite{BCHW-swapping} to bound it. In the first step, we will bound this quantity
by a function of the relative entropy distance under restrictive measurements introduced in \cite{Piani2009-relent}.

In \cite{vDamGrunwaldGill} a measure of nonlocality, based on the relative entropy, was introduced 
(analogous measure was also used to quantify contextuality \cite{MeasureContext}). It captures quantitatively
how 
``similar'' is a given probability
distribution to a local one. Given
a box $\mathcal{P} = P(ab|xy)$, where for fixed $x,y$ we have distribution $P_{xy}(ab|xy)$, its nonlocality is quantified by:
\be
\mathcal{N}(\mathcal{P}) = \sup_{\{p(x,y)\}} \inf_{P_L \in L} \sum_{x,y} p(x,y) D(P_{xy}(ab|xy) || P_L(ab|x,y))
\ee
where infinum in the above is taken over all boxes admitting a local model (belonging to set L) and $D(P||Q)$  
is the relative entropy between distributions $P$ and $Q$, $D(P||Q) \equiv \sum_i P(i) \log \frac{P(i)}{Q(i)}$.

We are interested in quantifying  how much nonlocality $\mathcal{N}$ one can obtain from $n$ copies of a given state $\rho_{AB}$, per number of copies, 
in the asymptotic limit, after processing it by LOCC.
{\definition For a bipartite state $\rho_{AB}$ its Asymptotic relative entropy of nonlocality, $R(\rho_{AB})$, is given by:
\be
R(\rho_{AB}) \equiv \varlimsup_{n \rightarrow \infty}{1\over n} \sup_{\Lambda \in LOCC} \sup_{\DE{M_{xy}}} \mathcal{N}(\{Tr M_{xy}\Lambda(\rho_{AB}^{\otimes n})\}),
\ee
where $\varlimsup$ denotes the supremum limit.
}

Now we want to set bounds for the nonlocality attainable in the asymptotic scenario.
To state the bound 
we will need a well known entanglement measure, called relative entropy of entanglement \cite{DefER}:
$
E_r(\rho) = \inf_{\sigma \in SEP} S(\rho||\sigma)
$,
where $S(\rho||\sigma) = Tr\rho\log \rho - Tr\rho\log \sigma$ is the quantum relative entropy, and infimum is taken over separable states.

Our main results state an upper bound on the asymptotic relative entropy of (hidden) nonlocality of a PPT bipartite quantum state by 
the relative entropy of the partially transposed state. To achieve these results, 
we first introduce another nonlocality measure, which is at the same time an entanglement monotone, denoted as $T^{\infty}$. 

{\definition  For a bipartite state $\rho_{AB}$, its restricted regularized relative entropy of nonlocality is given by:
\begin{align} 
T^{\infty}(\rho_{AB})  \equiv & \varlimsup_{n\rightarrow \infty} {1\over n} \sup_{\Lambda\in LOCC} \sup_{\{M_{xy}\}} \sup_{p(x,y)} \inf_{\sigma \in SEP}  \\
 \sum_{x,y} & p(x,y) D(\{Tr M_{xy}\Lambda(\rho^{\otimes n})\}||\{Tr M_{xy}\Lambda(\sigma^{\otimes n})\}). \nonumber
\end{align}
}

Note that the definition of $T^{\infty}$ originates from $R$ by relaxing the optimization over local boxes to an optimization over separable states 
and same local measurements. 

Now we are ready to state our main result.
{\theorem For any bipartite state it holds that
\be
R(\rho_{AB}) \leq T^{\infty}(\rho_{AB}) \leq E_r(\rho_{AB}). 
\label{eq:RT}
\ee
For $\rho_{AB}$ a PPT state, it holds that
\be
T^{\infty}(\rho_{AB}) \leq E_r(\rho_{AB}^{\Gamma}).
\ee
\label{thm:R-T}
}

Which leads to the corollary:
{\corollary For a bipartite PPT state $\rho_{AB}$ it holds that:
\be
R(\rho_{AB}) \leq \min\DE{E_r(\rho_{AB}), E_r(\rho_{AB}^{\Gamma})}.
\ee
\label{cor:postselection}
}

Since $E_r$ is \textit{asymptotically continuous} \cite{Synak05-asym}, the bound $R(\rho_{AB}) \leq E_r(\rho_{AB})$ is meaningful only when the state is close to
separable states under global operations. More important is the second bound which, as we show here, leads to non-trivial examples.

{\it Bound on post-selected nonlocality.---} We can also extend the results of the previous section to non-trace-preserving maps, \ie when the parties 
can perform a `filtering'
operation before the Bell test, the so called hidden nonlocality scenario \cite{Popescu1995}.
Popescu \cite{Popescu1995} showed that by performing a `filtering'
operation, and given that 
this operation succeeds, it is possible to obtain much larger violation of the CHSH game on the resulting state, not bounded as we claimed. 
However, we note that it is also important to take into account the probability of obtaining the  `filtered' result. 
For this reason, in order to quantify the efect of postselection, we propose to consider
a {\it asymptotic relative entropy of hidden-nonlocality}, $R_H(\rho_{AB})$, defined as follows:
\begin{align}
R_H(\rho_{AB})  & \equiv  \varlimsup_{n \rightarrow \infty}{1\over n} \\ 
&\sup_{\Lambda \in LOCC}  \sup_{\DE{M_{xy}}} \sup_{F_0} p^{F_0} \mathcal{N}(\{Tr M_{xy}F_0(\Lambda(\rho_{AB}^{\otimes n}))\}).\nonumber
\end{align}
Where a filtering process, $F_0$, takes state $\Lambda(\rho_{AB}^{\otimes n})$ 
to flag form, $F(\rho)=\sum_i \ketbra{i}{i}\otimes F_{i} \rho F_{i}^{\dagger}$, and later  erasures all other results except the  ``good'' 
one that leads to the highest violation of the Bell inequality. $p^{F_0}= Tr F_{0}\Lambda(\rho_{AB}^{\otimes n})F_0^{\dagger}$ is
the probability that the filter results in the desired outcome. 
We can have the same bound for $R_H$, as for $R$:

{\theorem For any bipartite state $\rho_{AB}$ it holds that
\be
R_H(\rho_{AB}) \leq T^{\infty}(\rho_{AB}) \leq E_r(\rho_{AB}).
\ee

For a bipartite PPT state $\rho_{AB}$ it holds that
\be
R_H(\rho_{AB}) \leq E_r(\rho_{AB}^{\Gamma}).
\ee
\label{ThmRH}
}

{\it Application.---} An application of the Corollary \ref{cor:postselection} follows from the fact, 
that $E_r$ is asymptotically continuous \cite{Synak05-asym}, hence generally, if
$\rho_{\epsilon} \in D(\epsilon)$, for $\epsilon < {1\over 2}$ we have:
\be
R(\rho_{\epsilon}) \leq 4\epsilon \log d + 2 h(\epsilon)
\label{eq:ergamma-bound}
\ee
where $h(p)=-p\log p -(1-p)\log (1-p)$ is the binary Shannon entropy, and $d$ is the dimension of the system (due to Theorem \ref{ThmRH}
the same bound holds for $R_H$). Hence, if $\epsilon$ decreases with $d$ faster 
than $1\over \log d$,  the asymptotic relative entropy of nonlocality vanishes with increasing dimension 
(e.g., the family of states shown 
in eq. (\ref{eq:ppt-1example}), have this property).

{\it On implications for cryptography.---} In \cite{HHHLO:unco-pbit, HLLO-PRL} it is shown that one can launch quantum key distribution (QKD) protocols based on shared private bits. Here we are interested in device independent QKD protocols based on a (non pure) private states.
To this end, a private bit should violate some Bell inequality.
Indeed any known DI QKD protocol \cite{BHK-di-security,Hanggi-Renner-Wolf,BCK-di-security} is based on some Bell inequality $\mathcal{S}$,
and admits some level of violation, say $\epsilon_v$, below which it aborts.
Now, due to Eqs. (\ref{eq:1-example}-\ref{eq:prop-example}) there are (approximate) private bits, which exhibit violation of inequality
$\mathcal{S}$ only up to $\epsilon' < \epsilon_v$, and hence will be aborted. This rules out such states from usage in this particular
DI QKD protocol. Moreover, every realization of DI QKD has inevitable errors due to decoherence. In such a case, the level of violation
$\epsilon'$ can be even below the precision of the experiment.
In \cite{karol-PhD} it is shown that in some cases one needs a number of private bits scaling with the dimension of the state
in order to discriminate it from separable states by some restricted class of operations including separable ones.
Hence, the limitation in the case of DI QKD reported above is in agreement with this result,
since any Bell operator $\mathbf{S}$ witnessing nonlocality is an entanglement witness.
An interesting question for further investigation is if there is a difference in terms of key rates between QKD and 
DI QKD protocols based on non-pure private states.

{\it Discussion.---} We have presented bounds on quantum nonlocality, both, in the single copy case
for arbitrary 
bipartite states, as well as in the asymptotic and hidden-nonlocality scenarios for states with positive partial transpose. 
Although we use partial transposition techniques, our method is based on the concept of
state discrimination via restricted classes of operations: LOCC operations and the separable ones, which is a well established
problem in quantum information theory. We believe that new, much tighter, upper bounds can be
found using powerful results from the latter subject.

As future directions, for the single copy scenario, instead of discrimination
from separable states, a refinement would be to consider the distance from states admitting local hidden-variable model,
e.g. the class of Werner states with local model \cite{Barrett-local}, which could possibly lead to tighter bounds. 
For the asymptotic and hidden
nonlocality scenarios, it would be interesting to extend the bound for the asymptotic relative entropy of (hidden) nonlocality to the case of NPT states. 
Also for these scenarios it would be worth finding new bounds for  states invariant under partial transposition (especially for 
the ones containing private key \cite{smallkey,enclanglement}).

It is worth noting, that our results are strongly related to the so called {\it Peres conjecture} \cite{Peres-conjecture}, 
recently disproved in \cite{Vertesi-Brunner-disproving}. Namely, we have asked a quantitative rephrasing of the 
original question posed by Asher Peres: how much one can violate a Bell inequality with PPT states? We have shown that,
for {\it certain} PPT states, the level of violation, both for single copy as well as in terms of the relative entropy of (hidden) nonlocality in the asymptotic cases, 
can be negligible. 
Notably, as we showed in the examples, even some states containing privacy admit such limited nonlocality content.

{\it Acknowledgements.---} 
We thank M. Horodecki, M.T. Quintino, A. Winter and R. Ramanathan for helpful discussions. And  R. Garc\'ia-Patr\'on for valuable comments.
We also thank anonymous referees of the \textit{Theory of Quantum Computation, Communication and Cryptography} conference (TQC 2015) for pointing out 
to us the H\"{o}lder's inequality in order to extend Theorem 1 to Bell inequalities with negative coefficients.
K.H. and G.M. acknowledge Polish MNiSW Ideas-Plus Grant IdP2011000361, ERC Advanced Grant QOLAPS and 
National Science Centre project Maestro DEC-2011/02/A/ST2/00305. 
G.M.  is supported by Funda\c{c}\~{a}o de Amparo \`{a} Pesquisa do Estado de Minas Gerais.

\bibliography{references}

\onecolumngrid
\appendix*
\section*{Supplemental Material}

Here we present the detailed proofs of the results stated in the main text. We follow the same notation and all the numberings of equations and statements
refer to the main text.


\section{Bounds for Bell inequalities.}

We start by giving upper bounds on the maximum violation of a Bell inequality achieved by a quantum state in the single copy scenario. 
In the main text we have proved Theorem 1 for positive coefficients Bell inequality, here we present the proof for general coefficients:

{\it Proof of Theorem 1}.
Theorem 1 for general coefficients follows directly by the relations:
\begin{align}
|\mathbf{S}(\rho) - \mathbf{S}(\sigma)| = & |\sum_{a,b,x,y} Tr s^{a,b}_{x,y} A_{a|x}\otimes B_{b|y} (\rho - \sigma)|  \nonumber \\
 =&|\sum_{a,b,x,y} Tr s^{a,b}_{x,y} A_{a|x}\otimes (B_{b|y})^T (\rho - \sigma)^{\Gamma}|  \nonumber\\
=& | Tr\, \mathbf{S}^{\Gamma}(\rho^{\Gamma} -\sigma^{\Gamma}) | \label{eq:infty-norm}\\
\leq & Tr\, |\mathbf{S}^{\Gamma}(\rho^{\Gamma} -\sigma^{\Gamma}) | \nonumber \\
\leq &||\mathbf{S}^{\Gamma}||_{\infty}||\rho^{\Gamma} - \label{eq:positive} \sigma^{\Gamma}||.\nonumber
\end{align}
where in the fourth step we use the triangle inequality, and the last step follows from H\"{o}lder's inequalities for $p-$norms.
\qed

\begin{corone} For any bipartite Bell expression $\mathcal{S}$, and state $\rho$, it holds that:
\be
Q_{\mathcal{S}}(\rho) \leq C(\mathcal{S}) + Q(\mathcal{S})\inf_{\sigma \in SEP} ||\rho^{\Gamma}  - \sigma^{\Gamma}|| . \nonumber
\ee
\end{corone}

{\it Proof of Corollary 1}.
First note that by substituting any separable state  $\sigma$ in eq. (1), and using the fact that
$\mathbf{S}(\sigma_{AB}) \leq C(\mathcal{S}) \;\; \forall \; \sigma_{AB} \in SEP$
we have: 
\be   
\mathbf{S}(\rho) \leq C(\mathcal{S}) + ||\mathbf{S}^{\Gamma}||_{\infty}\inf_{\sigma \in SEP} ||\rho^{\Gamma}  - \sigma^{\Gamma}||.
\ee
Now taking supremum over POVMS $\{A_{a|x}\ot B_{b|y}\}$
on both sides we have the desired result.
\qed

Based on \cite{karol-PhD,BCHW-swapping}, we have by Corollary 1 an immediate observation that certain private states have a limited possibility of violating 
a Bell inequality.

{\observation For any bipartite Bell inequality $\mathcal{S}$, if the states $\sqrt{XX^{\dagger}}$ and $\sqrt{X^{\dagger}X}$ are separable, 
then a private state $\gamma_X$, described by $X$ according to eq. (7), satisfies:
\be
Q(\gamma_X) \leq C(\mathcal{S}) + Q(\mathcal{S})||X^{\Gamma}||.
\ee
While, as  shown before, $||X^{\Gamma}||$ can be vanishing exponentially fast in number of qubits that composes $\gamma_X$.}

\begin{prop1}
There exist bipartite states (see eq. \eqref{eq:rec-state-presented} bellow) $\rho \in B(\ccal^2\otimes\ccal^2 \otimes (\ccal^{d^k}\otimes\ccal^{d^k})^{\otimes m})$ with $d=m^2$, $k=m$ satisfying
$K_D(\rho^{\Gamma}\otimes\rho) \rightarrow 1$ with increasing $m$, such that: 
\be
Q_{\mathcal{S}}(\rho\otimes\rho^{\Gamma})\leq C(\mathcal{S}) +{ Q(\mathcal{S})\over 2^{m-1}}. \nonumber
\ee
\end{prop1}

{\it Proof of Proposition 1}. Consider $\rho$ defined in eq. (149) of \cite{BCHW-swapping}. 
It has the property that its distillable key is almost 1, and there is a separable state $\sigma_{\rho}$ 
such that $||\rho^{\Gamma}- \sigma_{\rho}^{\Gamma}||\leq p$, with 
$p = {({1\over 2} - q)^m\over 2q^m+ 2({1\over 2} - q)^m}$ and $q ={1\over 3}$. Note that $p \leq {1\over 2^m}$ for natural $m$. Knowing this, we
bound the violation achieved by the above states in two steps.
We first apply Theorem 1 to  state $\rho\otimes\rho^{\Gamma}$, with $\sigma = \sigma_{\rho}\otimes \rho^{\Gamma}$, in order to obtain:
\be
\mathbf{S}(\rho\otimes\rho^{\Gamma}) \leq \mathbf{S}(\sigma_{\rho}\otimes\rho^{\Gamma}) + Q(\mathcal{S})||(\rho\otimes\rho^{\Gamma})^{\Gamma} - (\sigma_{\rho}\otimes\rho^{\Gamma})^{\Gamma}||,
\ee
which in turn is bounded by 
\be
\mathbf{S}(\sigma_{\rho}\otimes\rho^{\Gamma}) + Q(\mathcal{S})||\rho^{\Gamma} - \sigma_{\rho}^{\Gamma}||.
\ee
We now use the fact that $\sigma_{\rho}$ is separable, and that we can write $\rho^{\Gamma} = (1-r) \rho_{sep} + r\rho_{nsep}$, 
with $\rho_{sep} \in SEP$ and $r \leq p$. By linearity of trace we obtain:
\be
\mathbf{S}(\sigma_{\rho}\otimes\rho^{\Gamma}) = (1-r) \mathbf{S}(\sigma_{\rho}\otimes\rho_{sep}) + r \mathbf{S}(\sigma_{\rho}\otimes\rho_{nsep}).
\ee
First term of RHS is bounded by $C(\mathcal{S})$, as the state $\sigma_{\rho}\otimes \rho_{sep}$ is separable. The second term is bounded
by $Q(\sigma_{\rho}\otimes\rho_{nsep})$, which is in fact equal to $Q(\rho_{nsep})$.

This lead us to the following bound:
\ben
\mathbf{S}(\rho\otimes \rho^{\Gamma}) \leq C(\mathcal{S}) +   pQ(\rho_{nsep})+  p Q(\mathcal{S})    \leq C(\mathcal{S})+ 2pQ(\mathcal{S}) \leq C(\mathcal{S}) + {Q(\mathcal{S})\over 2^{m-1}},  
\een
which proves the result.  

\qed

An example of states
satisfying Proposition 1 is the family $\hat{\rho}_{p,d,k,m}$ on $B\left(\ccal^2\ot\ccal^2\ot(\ccal^{d^k}\ot\ccal^{d^k})^{\ot m}\right)$ 
\cite{pptkey} (see \cite{BCHW-swapping}).
Their matrix form is given below, up to the normalization factor $N_m = 2(p^m)+2({1\over 2} -p)^m$:
\be 
\left[\begin{array}{cccc}
[p({\tau_1+\tau_2\over 2})]^{\ot m} &0&0&[p({\tau_1-\tau_2\over 2})]^{\ot m} \\
0& [({1\over 2}-p)\tau_2]^{\ot m}&0&0 \\
0&0&[({1\over 2}-p)\tau_2]^{\ot m}& 0\\
{[p({\tau_1-\tau_2\over 2})]}^{\ot m}&0&0&{[p({\tau_1+\tau_2\over 2})]}^{\ot m}\\
\end{array}
\right].
\label{eq:rec-state-presented}
\ee
$\tau_1 = ({\rho_a + \rho_s\over 2})^{\ot k}$ and $\tau_2=(\rho_s)^{\ot k}$, while $\rho_s$ and $\rho_a$ are the $d$-dimensional symmetric and antisymmetric Werner state, respectively.

\section{ Bound on asymptotic nonlocality}

To treat the asymptotic scenario we introduce the restricted regularized relative entropy of nonlocality.
This quantity is an entanglement measure, and is related to the relative entropy of nonlocality:

\ben 
T^{\infty}(\rho_{AB}) \equiv \varlimsup_{n\rightarrow \infty} {1\over n} \sup_{\Lambda\in LOCC} \sup_{\{M_{xy}\}}   \sup_{p(x,y)} \inf_{\sigma \in SEP}  \sum_{x,y} p(x,y) D(\{Tr M_{xy}\Lambda(\rho^{\otimes n})\}||\{Tr M_{xy}\Lambda(\sigma^{\otimes n})\}).
\een

In the expression $D(\{Tr M_{xy}\Lambda(\rho)\}||\{Tr M_{xy}\Lambda(\sigma)\})$, we can treat $\{Tr M_{xy}\Lambda(\rho)\}$ 
 as a diagonal matrix
 with elements given by the probability distribution $P_{xy}(ab|xy)$, and then $D(\cdot||\cdot)$ is the quantum relative entropy. 
A similar quantity has been introduced by Piani in \cite{Piani2009-relent}. This function is easier to deal with than $R(\rho)$.
Note that the definition of $T^{\infty}$ originates from $R$ by relaxing the optimization over local boxes to an optimization over separable states 
and same local measurements, other relaxations can be defined similarly giving rise to upper bounds of independent interests. 

Recalling the definitions introduced in the main text, we use the relative entropy as a measure of nonlocality
\be
\mathcal{N}(\mathcal{P}) = \sup_{\{p(x,y)\}} \inf_{P_L \in L} \sum_{x,y} p(x,y) D(P_{xy}(ab|xy) || P_L(ab|x,y)),
\ee
and for the asymptotic scenario we define the asymptotic relative entropy of nonlocality:
\be
R(\rho_{AB}) \equiv \varlimsup_{n \rightarrow \infty}{1\over n} \sup_{\Lambda \in LOCC} \sup_{\DE{M_{xy}}} \mathcal{N}(\{Tr M_{xy}\Lambda(\rho_{AB}^{\otimes n})\}).
\ee

In all the following proofs we consider optimization over the probability distribution of the inputs $\{p(x,y)\}$, but one can also restrict to the uniform case
 \cite{vDamGrunwaldGill} and all the results follow in the same way.

We are now ready to prove Theorems 2.

\begin{thm2} For any bipartite state it holds that
\be
R(\rho_{AB}) \leq T^{\infty}(\rho_{AB})\leq E_r(\rho_{AB}). 
\label{eq:RT2}
\ee
For $\rho_{AB}$ a PPT state, it holds that
\be\label{T-Er}
T^{\infty}(\rho_{AB}) \leq E_r(\rho_{AB}^{\Gamma}).
\ee
\end{thm2}

{\it Proof of Theorem 2}. We first prove that $R\leq T^{\infty}$.  
In the first step let us note that $\Lambda(\sigma^{\otimes n})$ is a separable state, since $\Lambda$ is an LOCC operation. Hence,
if we place infimum over all separable states $\sigma$  instead of that of the form $\Lambda(\sigma^{\otimes n})$ in 
definition of $T^{\infty}$, we may only decrease the quantity. Second, we observe that instead of obtaining the local quantum 
box via the same POVMs, $\DE{M_{xy}}$, as for $\Lambda(\rho^{\otimes n})$, we can place also infimum over all $\tilde{M}_{xy}$ acting on $\sigma$,
which also can only lower the quantity. In the last step we observe that the set of such obtained quantum boxes is included 
in the set of the local ones, hence we can place infumum over the latter instead, which proves the desired result.

Now we prove the relation $T^{\infty}(\rho) \leq E_r(\rho^{\Gamma})$ for PPT states. The proof of $T^{\infty}(\rho) \leq E_r(\rho)$ follows
in the same way, without use of partial transposition. To prove that $T^{\infty}(\rho) \leq E_r(\rho^{\Gamma})$ note that,  
since $\Lambda \in LOCC$, it has a separable representation: $\Lambda(\rho) = \sum_{i,j} C_i\otimes D_j(\rho)C_i^{\dagger} \otimes D_j^{\dagger}$.
Using properties of trace and the separable representation we focus now on the term:

\ben
D(\{Tr \sum_{ij}C_i^{\dagger}A_{a|x}C_i\otimes D_j^{\dagger}B_{b|y}D_j\rho^{\otimes n}\} ||\{Tr\sum_{ij}C_i^{\dagger}A_{a|x}C_i\otimes D_j^{\dagger}B_{b|y}D_j \sigma'^{\otimes n} \}).
\een
Applying to both its components the identity $Tr XY = Tr X^{\Gamma}Y^{\Gamma}$, we have
\ben
D(\{Tr \sum_{ij}C_i^{\dagger}A_{a|x}C_i\otimes (D_j^{*})^{\dagger}B_{b|y}^{T}D_j^{*}(\rho^{\Gamma})^{\otimes n})\}||\{Tr\sum_{ij}C_i^{\dagger}A_{a|x}C_i\otimes (D_j^{*})^{\dagger}B_{b|y}^{T}D_j^{*} (\sigma'^{\Gamma})^{\otimes n}\}) ,
\een
which implies that $T^{\infty}$
can be written as:
\ben
\lim_{n\rightarrow \infty} {1\over n} \sup_{\Lambda' \in LOCC} \sup_{\DE{M'_{xy}}} \sup_{\DE{p(x,y)}} \inf_{\sigma' \in SEP}   \sum_{x,y} p(x,y) D(\{Tr M'_{xy}\Lambda'({\rho^{\Gamma}}^{\otimes n})\}||\{Tr M'_{xy}\Lambda'({\sigma'^{\Gamma}}^{\otimes n})\}) 
\label{eq:Tb1}
\een
with $\Lambda'$ being a new separable operation with $D_j$ operators complex conjugated,  and  $M'$ being a new set of POVMs with $B_{b|y}$ transposed.
Now, since $\sigma'^{\Gamma}$ is also a separable state, and by the fact that the relative entropy is non-increasing under completely 
positive trace-preserving maps, we have:
\ben
\inf_{\sigma' \in SEP}\sum_{x,y}p(x,y)  D(\{Tr M'_{xy}\Lambda'({\rho^{\Gamma}}^{\otimes n})\}||\{Tr M'_{xy}\Lambda'({\sigma'^{\Gamma}}^{\otimes n})\})   \leq \inf_{\sigma \in SEP} \sum_{x,y} p(x,y) D({\rho^{\Gamma}}^{\otimes n}||\sigma^{\otimes n}).
\label{eq:Tb2}
\een
We finally use the identity $D({\rho^{\Gamma}}^{\otimes n}||\sigma^{\otimes n}) = n D(\rho^{\Gamma}||\sigma)$. Since the latter term does not depend on $\Lambda$ and $M$ as well as $p(x,y)$, 
and the number of copies $n$ cancels with the regularization term ${1\over n}$, we obtain via (\ref{eq:Tb1}) and (\ref{eq:Tb2}) the bound:
\be
T^{\infty}(\rho) \leq \inf_{\sigma \in SEP} D(\rho^{\Gamma}||\sigma) \equiv E_r(\rho^{\Gamma}).
\ee
\qed

\section{Bound on post-selected nonlocality}

We can also treat the case where a filtering (non-trace-preserving) operation is performed before the Bell test. To quantify the nonlocality achieved in this case we 
define the \textit{asymptotic relative entropy of hidden-nonlocality}:

\begin{align}
R_H(\rho_{AB})  \equiv & \varlimsup_{n \rightarrow \infty}{1\over n} \sup_{\Lambda \in LOCC}  \sup_{\DE{M_{xy}}} \sup_{F_0} p^{F_0} \mathcal{N}(\{Tr M_{xy}\Lambda(F_0(\rho_{AB}^{\otimes n}))\}),
\end{align}
where a filtering process, $F_0$, takes state $\Lambda(\rho_{AB}^{\otimes n})$ 
to flag form, $F(\rho)=\sum_i \ketbra{i}{i}\otimes F_{i} \rho F_{i}^{\dagger}$, and later  erasures all other results except the  ``good'' 
one that leads to the highest violation of the Bell inequality. $p^{F_0}= Tr F_{0}\Lambda(\rho_{AB}^{\otimes n})F_0^{\dagger}$ is
the probability that the filter results in the desired outcome.

Analogously to Theorem 2, for the hidden-nonlocality scenario, we have the following result:

\begin{thm3}
 For any bipartite state $\rho_{AB}$ it holds that
\be
R_H(\rho_{AB}) \leq T^{\infty}(\rho_{AB})\leq E_r(\rho_{AB}).
\ee

For a bipartite PPT state $\rho_{AB}$ it holds that
\be\label{RH-Er}
R_H(\rho_{AB}) \leq E_r(\rho_{AB}^{\Gamma}).
\ee\
\end{thm3}

{\it Proof of Theorem 3}. We just have to show that $R_H(\rho_{AB}) \leq T^{\infty}(\rho_{AB})$ and \eqref{RH-Er} follows from \eqref{T-Er}.

Let us consider:

\begin{align}
T_{n}(\rho_f) \equiv {1\over n} \sup_{\Lambda\in LOCC} \sup_{\{M_{xy}\}}   \sup_{p(x,y)} \inf_{\sigma \in SEP}\sum_{x,y} p(x,y) D(\{Tr M_{xy}\Lambda(\rho^{\otimes n})\}||\{Tr M_{xy}\Lambda(\sigma^{\otimes n})\}).
\end{align}

Now let us restrict to a map $\Lambda$ of the form $\Lambda= F \circ \Lambda_0$, where $\Lambda_0$ is an arbitrary LOCC operation that acts on 
$n$ copies of the system, and is followed by measurement $F$,
\ie $\Lambda(\rho^{\otimes n})=\sum_i p^{F_i} \ketbra{i}{i} \otimes F_i (\Lambda_0(\rho^{\otimes n})) F_i^{\dagger}$.   

This restriction just decrease the RHS, so we have
\begin{align}
 T_n(\rho^{\otimes n})  \geq  {1\over n}& \sup_{\Lambda_0\in LOCC} \sup_{F}  \sup_{\{M_{xy}\}}   \sup_{p(x,y)} \inf_{\sigma \in SEP} \\
 & p^{F_0} \sum_{x,y} p(x,y) D(\{Tr M_{xy}\sum_i p^{F_i} \ketbra{i}{i} \otimes F_i (\Lambda_0(\rho^{\otimes n})) F_i^{\dagger}||\{Tr M_{xy}\sum_i q^{F_i} \ketbra{i}{i} \otimes F_i (\Lambda_0(\sigma^{\otimes n})) F_i^{\dagger}),\nonumber
\end{align}
where $p^{F_i}= Tr F_i (\Lambda_0(\rho^{\otimes n})) F_i^{\dagger}$ and $q^{F_i}= Tr F_i (\Lambda_0(\sigma^{\otimes n})) F_i^{\dagger}$.

Using the following property of relative entropy \cite{Piani2009-relent}:
\be
D\de{\sum_i p_i \rho_i \otimes \ketbra{i}{i} \middle| \middle|\sum_i q_i \sigma_i \otimes \ketbra{i}{i}}=\sum_i p_i D(\rho_i||\sigma_i)+D(p||q),
\ee
we obtain
\begin{align}
T_n(\rho^{\otimes n})  \geq {1\over n} \sup_{\Lambda_0\in LOCC} \sup_{F_0}  \sup_{\{M_{xy}\}}   \sup_{p(x,y)} \inf_{\sigma \in SEP} p^{F_0} \sum_{x,y} p(x,y) D(\{Tr M_{xy}F_0(\Lambda_0(\rho^{\otimes n}))F_0^{\dagger}\}||\{Tr M_{xy}F_0(\Lambda_0(\sigma^{\otimes n}))F_0^{\dagger}\}).
\end{align}
 where we have dropped the terms $D(p||q)\geq 0$, $\sum_{F_i\neq F_0} p^{F_i} D(\rho_i||\sigma_i) \geq 0$.

Now note that the RHS is an upper bound for $R_H$ and then we have the desired result.

\qed

\end{document}